\documentclass[
superscriptaddress,
amsmath,amssymb,
twocolumn,
prl,
floatfix,
]{revtex4}

\usepackage{graphicx}
\usepackage{amsmath}
\usepackage{amssymb}
\usepackage{color}
\usepackage[colorlinks=true,citecolor=blue,linkcolor=magenta]{hyperref}
\usepackage[usenames,dvipsnames]{xcolor}

\newcommand{\rs}{\rm \scriptscriptstyle}

\begin{document}

\title{
Dipole Interaction Mediated Laser Cooling of Polar Molecules to Ultra-cold
Temperatures
}

\author{S.~D. Huber}
\affiliation{Department of Condensed Matter Physics, 
The Weizmann Institute of Science, 
Rehovot, 76100, Israel}

\author{H. P. B\"uchler}
\affiliation{Institute for Theoretical Physics III, 
University of Stuttgart, Germany}

\date{\today}

\begin{abstract}
We present a method to design a finite decay rate for excited rotational
states in polar molecules. The setup is based on a hybrid system of polar
molecules with atoms driven into a Rydberg state. The atoms and molecules
are coupled via the strong dipolar exchange interaction between two rotation
levels of the polar molecule and two Rydberg states. Such a controllable
decay rate opens the way to optically pump the hyperfine levels of polar
molecules and it enables the application of conventional laser cooling
techniques for cooling polar molecules into quantum degeneracy.
\end{abstract}

\maketitle

The fast decay of excited electronic state of atoms is a paradigm  of
interactions between light and matter. This decay is the main building
block for several technological applications like laser cooling or optical
pumping. In turn,  these technologies open the way to the potential
implementation of quantum simulators and scalable quantum computers. In
contrast to the atomic case, the decay of electronically excited states
in polar molecules generally ends up in an unwanted excited rotational or
vibrational internal state. This proves to be a major obstacle in combining
the tools for controlling and cooling atoms with the strong electronic
dipole-dipole interactions present in polar molecules.  In this letter,
we demonstrate the possibility to design an adjustable finite life time
for an excited {\em rotational level} in a polar molecule, which overcomes
the obstacles on the way towards optical pumping and laser cooling of
polar molecules.

Several alternative methods are experimentally developed  in order to
cool polar molecules which reside in the rotational and vibrational
ground state to cold temperatures. For example, buffer gas cooling
\cite{buuren09} and stark deceleration \cite{meerakker08}; see
Ref.~\cite{doyle04,polarmoleculebook} for a review.  However, currently
the most successful approach is to use ultra cold clouds of atoms from
which molecules are created via photo association or coherent formation
\cite{wang04,sage05,deiglmayr08}. This technique has recently been
perfectionized to create molecules in a well defined hyperfine state
\cite{ni08,ospelkaus10}. Removing the remaining kinetic energy turns out
to be extremely challenging due to the complex internal structure of
the polar molecules \cite{shuman10}. Given these complications it may
be profitable to use a hybrid setup of dipolar molecules with another
system. A gas of Rydberg atoms is a promising candidate as they have
both a large dipole moment and can be cooled to ultra cold temperatures
\cite{tong04,singer04,heidemann07,vogt06}.

\begin{figure}[htp]
\includegraphics[width= 1\columnwidth]{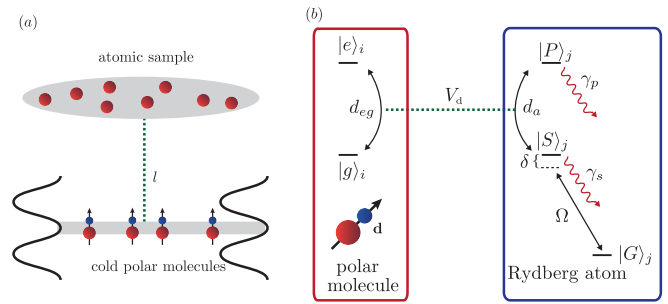}
\caption{
(a) Hybrid system: trapped polar molecules are in proximity to a
cloud of a cold atomic gas with a separation $l$. (b) Relevant level
structure:  two rotational states for the polar molecule $|e\rangle_{i}$
and $|g\rangle_{i}$ are coupled via dipole-dipole interaction to two
Rydberg levels $|S\rangle_{j}$ and $|P\rangle_{j}$ of an atom. In addition,
the atom is driven from the ground state $|G\rangle_{j}$ into a Rydberg
level. The rotational level $|e\rangle_{i}$ acquires a finite decay time
due to the resonant coupling and the finite life time of Rydberg levels.
}
\label{fig1} 
\end{figure}

In this letter, we demonstrate that a hybrid system of polar molecules
with Rydberg atoms \cite{zoller} allows for the design of a fast and
controllable decay for an internal rotational state.  The main idea
is based on the resonant and strong dipole-dipole interaction between
two rotational states and two Rydberg levels: The polar molecules make a
transition into the ground state, while at the same time the exchange of a
virtual microwave photon excites the Rydberg atom into a higher level, see
Fig.~\ref{fig1}. Then, the spontaneous decay of the Rydberg level induces a
finite effective decay rate for the rotational level of the polar molecules.

We show that such a decay opens up the way for optical pumping of hyperfine
levels and laser cooling of polar molecules using standard techniques.
Our setup has two main advantages. First, the decay rate is tunable to
arbitrary low values and, second, the decay  takes place with very low
recoil momentum. Consequently, the Doppler temperature is much lower than
in their atomic counter parts and allows one to cool polar molecules into a
regime where interactions and particle statistics play in important role.
We demonstrate that the method exhibits the potential for laser cooling
of strongly interacting many-body system into strongly correlated phases,
such as the crystalline phase of polar molecules. We envisage, that our
scheme is most suitable for polar molecules created by a coherent formation
\cite{ni08,ospelkaus10}, where unpaired atoms are naturally present,
and the polar molecules are already at rather low temperatures.

We start with the description of the setup for the engineering of a
controlled decay rate for an internal rotational state of polar molecules.
The decay takes place from the excited state denoted as $|e\rangle_{i}$
into the ground state $|g\rangle_{i}$ for the $i^{\rm th}$ polar molecule,
see Fig.~\ref{fig1}. The transition between the two states is characterized
by a dipole matrix element  ${\bf d}_{eg} = \langle g| {\bf d} |e \rangle$
with ${\bf d}$ the dipole operator. The internal structure of the polar
molecule is described by an effective spin $1/2$ system with the Hamiltonian
$H^{(i)}_{m}  =  \hbar \omega_{0}  \sigma_{i}^{z}/2$.  The energy difference
between the excited and the ground state is   $\hbar \omega_{0}$, and
its typical value is in the range of the rotational splitting $2 B$.
The polar molecule is coupled via the dipole-dipole interaction to the
Rydberg levels of a close by atom. The relevant levels for the atom are the
ground state $|G\rangle_{j}$ and the two Rydberg states $|S\rangle_{j}$
and $|P\rangle_{j}$. The ground state $|G\rangle_{j}$ is coupled via
a driving laser field to the Rydberg state $|S\rangle_{j}$. Within the
rotating frame and using the rotating wave approximation, the Hamiltonian
for the atom reduces to
\begin{equation}
  H^{(j)}_{a} =   
  \sum_{\alpha\in\{G,S,P\}} E_{\alpha} |\alpha\rangle \langle \alpha|_{j}    
  + \frac{\Omega}{2}
   \Big(   |G\rangle \langle S|_{j} + |S\rangle \langle G|_{j}\Big), 
\end{equation}
where $\Omega$ is the Rabi frequency of the driving laser field with the
detuning $\hbar \delta= E_{S}-E_{G}$. In addition, the Rydberg states
are characterized by a dipole transition matrix element ${\bf d}_{a}=
\langle S| {\bf d}|P\rangle$.

The dipole-dipole interaction gives rise to a resonant exchange interaction
between the polar molecules and the Rydberg levels $|e\rangle_{i}
|S\rangle_{j} \rightarrow |g\rangle_{i} | P\rangle_{j}$.  The setup
is designed such that the exchange takes place near resonant with the
detuning $\hbar \Delta= E_{P}-E_{S}- \hbar \omega_{0}$. The interaction
between the atom and the molecule reduces to
\begin{equation}
   H_{c}= \sum_{i,j} V_{d}({\bf r}_{i}-{\bf R}_{j})\left[S^{-}_{i} 
   T^{+}_{j}+ S^{+}_{i} T^{-}_{j}\right],
\end{equation}
where $V_{d}({\bf r}) = {\bf d}_{eg}\cdot {\bf d}_{a}/r^3 - 3 ({\bf d}_{eg}
\cdot{\bf r})({\bf d}_{a} \cdot{\bf r})/r^5$ denotes the dipole-dipole
interaction with ${\bf r}_{i}$ the positions of the polar molecules and
${\bf R}_{j}$ the coordinates of the atoms, respectively.  The operators
$S^{-}_{i} = |g\rangle\langle e|_{i}$ account for the transition within
the polar molecule, while   $T^{+}_{j} = |P\rangle \langle S|_{j}$   is
the corresponding operator acting on the Rydberg levels.

The excited electronic states of the atom $|S\rangle_{j}$ and $|P\rangle_{j}$
are characterized by a finite life time with the decay rates $\gamma_{\rs
S}$  and $\gamma_{\rs P}$, respectively. The time evolution of the full
system is then described by the master equation for the density matrix $\rho$
\begin{equation}
   \partial_{t} \rho = \frac{i}{\hbar} \left[H, \rho \right] 
   + \!\!\!\!\!\! \sum_{\alpha \in \{S,P\}} \!\!\!\frac{\gamma_{\alpha}}{2} 
   \left[ 2 c_{\alpha} \rho c_{\alpha}^{\dag}
    - c^{\dag} _{\alpha}c_{\alpha} \rho
   - \rho c^{\dag}_{\alpha} c_{\alpha}\right], \label{masterequation}
\end{equation}
with the jump operators $c_{\alpha} = |G\rangle \langle \alpha|$ with
$\alpha \in \{S,P\}$ accounting for the decay from the  Rydberg states
into the ground state $|G\rangle_{j}$.  In addition, the Hamiltonian $H$
contains the polar molecules as well as the driven atomic system, i.e.,
$H = \sum_{i} H_{m}^{(i)} +\sum_{j} H_{a}^{(j)} + H_{c}$.

The master equation Eq.~(\ref{masterequation}) implies that any initial
state ends up with all polar molecules in the rotational ground state
$|g\rangle_{i}$, as can be seen by a numerical simulation of the time
evolution. For weak couplings, it follows that the dynamics is well
described by an effective decay rate for the excited rotational state of
the polar molecules, which reduces to
\begin{equation}
   \Gamma_{i} =  \sum_{j} \frac{\left|M_{i j}\right|^2}{\hbar^2} 
   \frac{\gamma_{\rs P}}{\gamma_{\rs P}^2/4+(\Delta -\delta)^2}
   \frac{\Omega^2}{\gamma_{\rs S}^2/4 + \delta^2} .
   \label{decayrate}
\end{equation}
The last factor in the product $p_{s}=\Omega^2/(\gamma_{s}^2/4 + \delta^2)
$ describes the probability for an atom to be in the excited Rydberg state
$|S\rangle_{j}$ under the external drive . The second factor accounts for
the finite line width of the Rydberg states and the first factor is the
coupling matrix elements determined by the dipole-dipole interaction
\begin{equation}
 |M_{i j}|^2 = \sum_{n'm'} 
 | \langle \psi_{i n'} \phi_{j m'} | 
  V({\bf r}_{i}- {\bf R}_{j})
  | \psi_{i n}, \phi_{j m}\rangle|^2.
  \label{eqn:spatial}
\end{equation}
Here, $\psi_{i n}$  ($\phi_{jm}$) are  the initial spatial wave function and
$\psi_{i n'}$  ($\phi_{j m'}$) the final spatial wave function  for the polar
molecule (atom), respectively.  The decay rate in Eq.~(\ref{decayrate})
is the main result of this paper. In the following we provide (i) a
discussion of Eq.~(\ref{decayrate}) and the limits of its applicability,
(ii) numerical values relevant for concrete experiments, and (iii) a
possible application in the from of Doppler cooling.

{\bf Discussion of Eq.~(\ref{decayrate}) --}  We start with analyzing
the matrix elements $M_{i j}$ for two different setups: First, we
consider molecules tightly confined in a three dimensional harmonic
trap with oscillator length $a_{0}$. If $a_0$ is much smaller than the
distance between the atoms and the polar molecules $l$, we can express
the dipole-dipole interaction in terms of the harmonic oscillator wave
functions. The excitation from the oscillator ground state is then
suppressed by $(a_{0}/l)^{2}$. Consequently, in leading order the polar
molecule remains within the spatial ground state. However, the resulting
average energy transfer by $2 \hbar^2 \pi^2/m l^2$ corresponds to a recoil
momentum of $q_{d} = 2 \pi \hbar/l$.

Alternatively, for polar molecules tightly confined in only two
dimensions, see Fig.~\ref{fig1}, a transfer of in-plane momentum ${\bf
K}$ can appear. Its contribution is described by the in-plane Fourier
transformation of the interaction potential $V_{d}(l,{\bf K})= \int
d {\bf r}_{\parallel} V_{d}(l, {\bf r}_{\parallel}) e^{i {\bf K} {\bf
r}_{\parallel}}$. It is peaked for $ K \sim 1/l$ and decays exponentially
for larger momenta. Consequently, also such a setup gives rise to a
characteristic momentum recoil $q_{d} = 2 \pi \hbar /l$.

The above analysis reveals an important aspect of our setup: The recoil of
the spontaneously emitted photon from the Rydberg state only acts on the
atom, while the recoil for the polar molecule $q_d$ in turn is completely
determined by the separation $l$ and can be strongly suppressed compared
to the photon recoil. Note, that this discussion for the recoil energy
is limited to the regime, where Fermi's Golden rule is applicable. Hence,
we need to asses the validity of Eq.~(\ref{decayrate}).

Let us return to the evaluation of  Eq.~(\ref{decayrate}). The decay
rate is enhanced by the number of atoms within range of the dipole-dipole
interaction, and can be written as $ \sum_{j} |M_{i j}|^2 = N_{i} |V_{d}({\bf
l})|^2$. The number of atoms within interaction range $N_{i}$ strongly
depends on the spatial distribution  of the atomic sample and reduces
to  $N_{i} = \int d{\bf R}   |V_{d}({\bf r}_{i} - {\bf R})|^2  \rho({\bf
R})$. Here, $\rho({\bf R})$ denotes the density of Rydberg atoms. For
weak Rabi frequencies $ \Omega/\delta \ll 1$ this density is determined
by the atomic density reduced by a factor $p_{s}$ accounting for the
probability to find an atom in the Rydberg state. For strong drives the
Rydberg-blockade phenomena starts to play a role and limits the Rydberg
density; see \cite{low09} for a detailed discussion. In either case,
this analysis provides us with an effective coupling strength $N_{i}
|V_{d}({\bf l})|^2$ between a single molecule and the atomic cloud.

We can now answer the question on the validity of Fermi's Golden rule
which is based on weak coupling. In the presence of the enhancement
by the different decay channels, this condition requires $\sqrt{N_{i}}
|V_{d}({\bf l})| \lesssim\gamma_{p} $. As a consequence, the induced decay
rate $\Gamma_i$ is limited by the decay rate of the Rydberg level. However,
the latter can be strongly enhanced and controlled by dressing the Rydberg
level with a lower lying electronic state \cite{zoller}. Denoting the decay
rate of the additional electronic state by $\gamma_{d}$, the far-detuned
coupling of the Rydberg level $|P\rangle_{j}$ with this electronic state
with Rabi frequency $\Omega_{d}$ and detuning $\Delta_{d}$ provides a total
decay rate for the dressed Rydberg level $\gamma'_{p} \approx \gamma_{p}
+ \gamma_{d}  \: \Omega_{d}^2/4\Delta_{d}^2$. When providing numerical
values for the obtainable decay rates for the rotational levels with take
this enhancement into account.

{\bf Numerical values --} In the following, we provide experimental
parameters for LiCs polar molecules with the atomic reservoir given by
Cs atoms excited into the Rydberg state. First, we ignore the weak hyper
fine interaction for the polar molecules,  and describe the characteristic
energy scale; the incorporation of the nuclear spin is then straightforward.
We choose for the excited states of the polar molecule the first excited
rotational states with angular momentum one $|e\rangle =|1,0\rangle$; here
$|j,m\rangle$ denote angular momentum eigenstates of the rotational degree
of freedom.  The dipole moment to the ground state reduces to $|{\bf d}_{eg}|
= d/\sqrt{ 3}$ with $d = 5.5 \: {\rm Debye} \approx 2.14 e a_{0}$ the
permanent electric dipole moment of the polar molecule.  With a rotational
splitting $B\approx 5.84 {\rm GHz}$, the transition $|1,0\rangle \rightarrow
|g\rangle=|0,0\rangle$ becomes nearly resonant for the Rydberg transition
between states with principle quantum number $n=69$ and (electronic)
angular momentum $|s\rangle \rightarrow |p_{3/2}\rangle$. According to
the well established quantum defect theory \cite{rydbergbook} they have an
energy difference of $\Delta  \approx 40 {\rm MHz}$ and a dipole moment $
|{\bf d_{a}}| \approx 10 500 \: {\rm Debye}$.  With these large dipole
moments the resonant dipole-dipole interaction reduces to  $V_{d}({\bf l})
\approx 1.7 \: {\rm kHz} $ even at a distance of $l=25 {\rm \mu m}$ between
the polar molecules and the Rydberg atoms (separated along the z-direction).

This energy scale should be compared with the decay rate from the Rydberg
state $\gamma_{\rs p}$, which is dominated by black body radiation and
reduces to $\gamma_{p}\sim 840 {\rm Hz}$. In the following, we will enhance
this decay rate  to $\gamma_{p}' \approx 0.3 \: {\rm MHz}$ as discussed
above. Consequently, we obtain a decay rate $\approx 39{\rm Hz} $ for each
atom excited into the Rydberg level $|S\rangle$.

Next, we need to estimate the enhancement $N_i$ due to the coupling to a
whole ensemble of Rydberg atoms.  The characteristic blockade radius is
in the range of $\approx 3 {\rm \mu m}$ \cite{low09}, and consequently,
the Rydberg density is limited by $\approx 10^{11} {\rm cm}^{-3}$;
this density is much lower than the characteristic atomic density for
a cold sample of Cs atoms. For an atomic sample within an oblate trap
with transverse confining along the $z$-axis with $10 {\rm \mu m}$,
the summation over the different Rydberg atoms provides an enhancement
$N_{i}\approx  52000$. Then, the condition $ \sqrt{N_{i}} V_{d}({\bf l})
\lesssim \gamma_{p}'$ is satisfied and the application of Fermi's Golden
rule for the estimation of the induced decay rate is valid. We finally
find a induced decay rate for the rotational level of the polar molecules
$\Gamma_{i} \approx 0.25 {\rm MHz}$.

Alternatively, it is experimentally interesting to work in a parameter
regime with strong electric fields in combination with a strong confinement
of the molecules into a two-dimensional setup \cite{buechler07}. However,
the strong electric field requires to work with much lower main principal
number $n$ of the Rydberg levels to prevent field ionization.

In this setup the condition of resonant dipole-dipole interaction is
conveniently obtained by selecting two Rydberg states which undergo a
true level crossing within the stark map, but are coupled by exchanging a
quantum of angular momentum. For Cs Rydberg atoms, the states with $n=17$
and  $|S\rangle = |d_{5/2},\pm 3/2\rangle$ and  $|P\rangle =|d_{5/2},\pm
5/2\rangle$ exhibit such a level crossing \cite{zimmerman79} at an electric
field $E \approx 3 {\rm kV/cm}$. Moreover, they allow for a resonant dipole
exchange with the two states $|e\rangle = |1,\pm\rangle$  and $|g\rangle =
|0,0\rangle$ of the polar molecule \cite{note}.  On a distance with $l = 3
{\rm \mu m}$ the dipole exchange energy reduces to $V_{d}({\bf l}) \approx
3100 {\rm Hz}$.  With the decay rate for the Rydberg atoms $\gamma_{s}
\approx \gamma_{p} \approx 72 {\rm kHz}$, the induced decay rate with
a single Rydberg atom reduces to $ \approx 530 {\rm Hz}$.  Again, this
decay rate is enhanced by the number independent decay channels $N_i$.
Assuming the atomic system confined into a 2D layer above the polar
molecules with a Rydberg density $\approx 10^{12} {\rm cm}^{-3}$, the
total decay rate reduces to $\Gamma_{i} \approx 11 {\rm k Hz}$.

{\bf Applications --} An immediate application capitalizing on the
finite life-time of a rotational level in polar molecules is the optical
pumping of the hyper fine structure. The main mechanism is the electric
quadrupole interaction, which couples nuclear spins to rotational
excitations, and has been successfully used for the coherent transfer of
population between different hyperfine states \cite{ni08,ospelkaus10}.
Note, that the optical pumping may give rise to coherence between
polar molecules separated on distances smaller than $l$; it is this
coherence, which allows to apply conventional laser cooling techniques
\cite{stenholm86,phillips98,lasercoolingbook} to temperatures below
quantum degeneracy.

In the following, we describe the peculiarities of our setup for  Doppler
cooling. However, we  would like to stress, that the induced decay rate can
also be used for more sophisticated cooling techniques, such as Sisyphus
cooling, which in general will give rise to more efficient cooling of
the system.  In order to achieve Doppler cooling, the transition into
the excited rotational state is driven via a Raman transition with Rabi
frequency $\Omega$, the detuning from resonance $\Delta$, and a momentum
transfer ${\bf q}$. In contrast to conventional laser cooling, here, the
spontaneous decay from the excited rotational state transfers a different
momentum to the polar molecule. Consequently, we have to distinguish
between two momenta. The momentum ${\bf q}$ absorbed in the Raman process
and the small momentum $q_d$ from the spontaneous decay; see above for a
discussion of this momentum.

The standard laser cooling techniques are characterized by two energy
scales: the higher energy/temperature  $T_{\rs c} =m \Gamma^2/(2 {\bf q}^2)$
determines the kinetic energy, where laser cooling  becomes effective;
here $\Gamma$ denotes the induced decay rate from the excited rotational
state. Second, the equilibrium state induced by the laser cooling is
characterized by the Doppler temperature $T_{d} = \hbar \Gamma$.

A major advantage of the present scheme is the possibility  to tune
the recoil momentum ${\bf q}$ independently of the decay rate $\Gamma$.
As a consequence, the capturing temperature $T_{\rs c}$ and the Doppler
temperature can be adjusted for optimal cooling.  The Doppler temperature
of the equilibrium state can be continuously decreased by reducing the
decay rate $\Gamma$. The lowest temperature is therefore only limited by
the small recoil momentum $q_d$ from the spontaneous decay, i.e., $T_{\rs
limit} = \hbar^2 q_{d}^2/2 m$.  This temperature is in general much lower
than the Doppler temperature for conventional laser cooling of alkali atoms
\cite{lasercoolingbook}, and the present scheme allows to laser cool polar
molecules to temperatures comparable to quantum degeneracy.

Note, that for conventional Doppler cooling, the re-absorption of
emitted photon provides a decrease in cooling efficiency for dense
atomic samples. The analogue effect in the present scheme would be the
re-excitation of a polar molecule by a resonant dipolar exchange. However,
this rate is strongly suppressed for a fast decay of the Rydberg level and
a proper balance between the number of atoms $N$ and polar molecules $M$
with the requirement $M \ll N $.

Close to quantum degeneracy, the statistics of the particles starts to play
an important role. The Doppler cooling scheme described above remains valid
also in this regime. This allows one in principal to cool directly into a
Fermi sea for fermions or a Bose-Einstein condensate for bosons. However,
the Doppler cooling scheme breaks down at low temperatures and high particle
densities, where  interactions between the molecules become relevant.

There are two main effects due to interactions: first, the interaction
potential between two molecules in state $|g\rangle$, and the interaction
potential between the states $|g\rangle$ and $|e\rangle$ are in general
different; especially, as for the two states $|g\rangle$ and $|e\rangle$
also a resonant exchange of a microwave photon can appear.  During the
Doppler cooling process, each quantum jump with the excitation of a polar
molecule into the higher rotational state provides a fluctuating force
to the surrounding polar molecules and  eventually provides a source
of heating.  Second, the difference of the interaction potentials leads
to a level broadening of the excited rotational state, and the desired
energy selectivity for the Doppler cooling breaks down as soon as the
level broadening reaches the induced decay rate.

However, one can envisage a strategy that overcomes this limitation and
potentially allows  cooling directly into a strongly interacting many-body
phase such as the crystalline state \cite{buechler07}. The trick is to
adiabatically eliminate the excited rotational states by irradiating a
far-detuned Raman laser. Consequently, the many-body states acquire a finite
lifetime inherited from the small admixture of the excited state. In the
following, we denote by $|\alpha\rangle$ and $|\beta \rangle$ eigenstates
of the fully interacting many-body system with energies $E_{\alpha}$
and $E_{\beta}$, respectively. Using Fermi's Golden rule, we obtain
transition  rates $\Gamma_{\alpha \rightarrow \beta}$ from the initial
state $|\alpha\rangle$ into the final state $|\beta \rangle$ due to the
coupling of the system to the bath
\begin{equation}
 \Gamma_{\alpha \rightarrow \beta} = 
 \frac{\Omega^2}{4 \Delta^2} |M_{\alpha \beta}|^2 
 \frac{\Gamma}{\left(\tilde{\Delta} - E_{\alpha}+
  E_{\beta} \right)^2 + \Gamma^2/4}.
\end{equation}
Here, $\Gamma$ denotes the decay rate of the bath, while the matrix elements
for the many-body system reduces to
\begin{equation}
 M_{\alpha \beta} = \int d{\bf x}  \: 
 \langle  \alpha |  \psi^{\dag}({\bf x}) \psi({\bf x})  
 e^{i {\bf q}{\bf x}} | \beta \rangle.
\end{equation}
It is important to point out, that the energy dependence is in analogy to
conventional laser cooling techniques, i.e., for $E_{\alpha}- E_{\beta}>0$
and blue detuning $\Delta >0$, the transitions $|\alpha\rangle \rightarrow
|\beta\rangle$ take place with a higher rate than the reversed transitions
$|\beta\rangle  \rightarrow |\alpha\rangle $, and consequently leads to
a cooling of the system.

{\bf Conclusion.} In summary, we proposed a scheme to induce a controlled
decay rate for rotational levels of polar molecules. We showed how one
can harness this finite lifetime to perform Doppler cooling into quantum
degeneracy. Finally, we commented on how to implement a direct cooling
into a strongly interacting many-body phase.

{\it Acknowledgment:} We thank T. Pfau and P. Zoller for fruitful
discussions. Support by the Deutsche Forschungsgemeinschaft (DFG) within SFB
/ TRR 21 and National Science Foundation under Grant No. NSF PHY05-51164
is acknowledged. SDH acknowledges support by the Swiss Society of Friends
of the Weizmann Institute of Science.

\end{document}